# Self-learning photonic signal processor with an optical neural network chip

Hailong Zhou[1†], Yuhe Zhao[1†], Xu Wang[1], Dingshan Gao[1], Jianji Dong[1★] and Xinliang Zhang[1★]

[1]Wuhan National Laboratory for Optoelectronics, School of Optical and Electronic Information, Huazhong University of Science and Technology, Wuhan 430074, China; [†]These authors contributed equally to this work. [★]e-mail: jjdong@hust.edu.cn; xlzhang@mail.hust.edu.cn.

**Photonic signal processing is essential in the optical communication and optical computing. Numerous photonic signal processors have been proposed, but most of them exhibit limited reconfigurability and automaticity. A feature of fully automatic implementation and intelligent response is highly desirable for the multipurpose photonic signal processors. Here, we report and experimentally demonstrate a fully self-learning and reconfigurable photonic signal processor based on an optical neural network chip. The proposed photonic signal processor is capable of performing various functions including multichannel optical switching, optical multiple-input-multiple-output descrambler and tunable optical filter. All the functions are achieved by complete self-learning. Our demonstration suggests great potential for chip-scale fully programmable optical signal processing with artificial intelligence.**

Photonic signal processing is widespread both in the optical communication and optical computing. Almost all aspects of signal processing functions were developed, such as multichannel optical switching[1-7], mode (de)multiplexer[8,9], optical filter[10-14], polarization management[15-17], waveguide crossing[18,19] and logic gates[20-22]. Plentiful photonic signal processing devices have been reported based on either discrete components or photonic integrated circuits[1,9,15,20-24]. Photonic signal processing devices based on discrete components are usually bulkier and less power efficient, whereas a photonic integrated signal processing chip has a much smaller footprint and a higher power efficiency. To date, various separated functional devices may make the system complex and high-cost. An effective solution is to implement low-cost and multipurpose signal processing in an optical network with a reconfigurable and integrated photonic signal processor. For example, a fully reconfigurable photonic integrated signal processor performed numerous types of optical computing based on an InP–InGaAsP material system[25]. Multipurpose silicon photonic signal processors were reported based on hexagonal waveguide mesh[13,26,27]. Tunable filters[10-14] and programmable nanophotonic quantum processors[28-34] were also addressed. A big challenge is how to dynamically configure these functions for reconfigurable photonic signal processors. Most of the reported photonic signal processors were configured by manual operation[1,2,12,14,25,27,35]. It is cumbersome, time consuming and becomes extremely difficult when the network is expanded to a larger scale. Alternatively, some self-configuring methods were proposed. For example, the multipurpose processors could be trained using the feedback from the built-in optical power monitors[36-39], whereas the internal monitors will introduce additional loss and the number of monitors will increase rapidly with the expansion of network, making both the electronic circuit and iterative algorithm quite complex. Besides, the inner structure of chips should be revealed in advance to the operators before any configuration of the system. The network could be alternatively trained by pre-calibration and subsequently analytic calculation of phase in each phase shifter[13,26,27,40]. However, it strongly depends on the fabrication tolerance of chip network and the detailed information about chip inner structure. It will be greatly complicated and difficult when the network is expanded to a larger one, since too much inner information is involved. Therefore, a fully automatic implementation of reconfigurable photonic signal processors capable of full self-learning without knowing the detailed information of chip inner structure is highly required. In recent years, the optical neural networks (ONNs) have been developed to replace the traditional central processing units to perform deep-learning tasks, where the electrical devices performed the control and feedback of the ONNs, and photonic devices performed the computing of ONNs with a faster speed and a lower cost[41-43]. The development of deep learning with photonic hardware promotes the possibility of a fully self-learning photonic signal processor, which can learn to achieve several functions without the detailed information of chip inner structure.

In this paper, we report a fully self-learning photonic signal processor based on an ONN chip without activation functions or with identity activation functions ( $f(z) = z$ ), which can perform three typical functions, including multichannel optical switching, optical multiple-input-multiple-output (MIMO) descrambler and tunable optical filter. All the functions are programmable by self-learning starting from blank. The training is accomplished using the numerical gradient descent algorithm modified from deep learning[41,42], which is applicable for a "black box" system. This demonstration is an important step towards the implementation of a multifunctional optical signal processor with artificial intelligence.

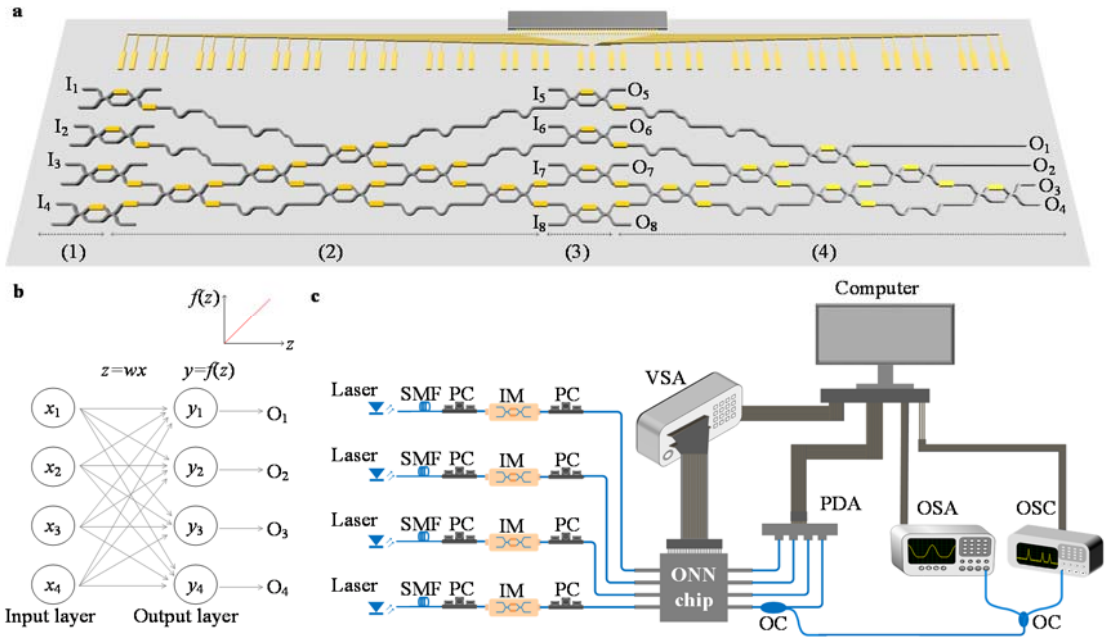

**Fig. 1| Optical signal processing circuits and schematic of the experimental set-up. a,** The detailed structure of ONN chip. The ONN chip contains 48 phase shifters and 20 Mach-Zehnder interferometers (MZIs) in total. It can be divided to four functional parts, which carry out the following: (1) switching the input channel; (2) generating the first unitary matrix; (3) generating the diagonal matrix; (4) generating the second unitary matrix. Parts (2), (3) and (4), as a whole, are used to implement an arbitrary matrix transformation. **b**, General ONN architecture composed of an input layer and an output layer. The activation functions of ONN are identical here. **c**, Schematic of the experimental set-up. Four tunable continuous wave lases with the same wavelength at 1550 nm are launched into four optical intensity modulators (IMs) respectively, by which four independent 10 Gbit/s non-return-to-zero (NRZ) signals are prepared. Before the IMs, four polarization controllers (PCs), connected by single mode fibers (SMFs), are used to tune polarization states to ensure the quality of NRZ signals. Then the four optical signals are launched into our optical neural network (ONN) chip through a V-groove fiber array (VGA). The polarizations of input/output light are optimized by in-line

polarization controllers (PCs). Beams emerging from the device are collected by another VGA and a 4-channel photodetector array (PDA). Part of beams in the any specified channel can be separated by fiber optic couplers (OCs) and then fed into an optical spectrum analyzer (OSA) or an optical oscilloscope (OSC) for further analysis. The ONN chip contains 48 phase shifters which are controlled by a voltage source array (VSA). The devices of VSA, PDA, OSA and OSC are all connected and controlled by the same computer.

The photonic signal processor, structured with an ONN shown in Fig. 1a[41,44,45], is fabricated on the commercial silicon-on-insulator (SOI) wafer (Supplementary Section 1). The activation functions of ONN are identity functions ($f(z) = z$) here. The ONN chip consists of four parts. Part (1) is constitutive of four Mach-Zehnder interferometers (MZIs), which are only in the "open" or "closed" state to switch the input channel. Parts (2) and (4) can both perform an arbitrary unitary matrix transformation (Supplementary Section 2). Part (3) can perform an arbitrary diagonal matrix transformation or be used to switch the input channel, as same as Part (1). Parts (2), (3) and (4), as a whole, are used to implement an arbitrary matrix transformation based on singular value decomposition[41,44]. The ONN chip contains 48 phase shifters and 20 MZIs totally. The general architecture of ONN is composed of an input layer and an output layer with identity activation functions, shown in Fig. 1b. The experimental set-up is depicted in Fig. 1c. Four independent 10 Gbit/s non-return-to-zero (NRZ) signals are loaded into four laser beams at 1550 nm with the optical intensity modulators (IMs). And then the beams are launched into our ONN chip through a V-groove fiber array (VGA). The polarizations of input/output light are optimized by in-line polarization controllers (PCs). Another VGA and a 4-channel photodetector array (PDA) are used to receive the output light from the chip. Part of beams in a specified channel are separated by fiber optic couplers (OCs) and then fed into an optical spectrum analyzer (OSA) or an optical oscilloscope (OSC) for further analysis. All the phase shifters in the ONN chip are driven by a voltage source array (VSA). All the monitoring instruments and voltage sources are connected and controlled by the same computer. Based on this set-up, multiple photonic signal processing functions can be implemented by self-learning. According to the different purposes of signal processors, a suitable and special cost function (CF) should be first defined for a successful training (Supplementary Section 3). Then the only learning target is to make the defined CF best using the numerical gradient descent algorithm. In the following, the ONN chip is reconfigured to achieve three different functions with self-learning.

**Multichannel optical switching**

An $n$-channel optical switching is a standard matrix transformation given by (referred to Supplementary Section 3)

$$M = \begin{bmatrix} M_1 & M_2 & \ldots & M_n \end{bmatrix}, \tag{1}$$

where $M_1, M_2, \ldots, M_n$ are the realignment of standard unit vectors in $n$ dimensional Hilbert space. $M_j$ is the $j$th column vector of $M$, representing the power distribution at all the output ports when only the input Port $j$ is switched on. The traditional optical link switching was accomplished by manual adjustment or semi-automaticity[1,5,7], here we exploit the self-learning method to implement the optical link switching automatically. In order to implement the optical switching irrelevant to the input powers, we define the CF as the correlation of experimental and theoretical vectors (see Supplementary Section 3)

$$\mathrm{CF} = \prod_{j=1}^{n} \mathrm{Corr}(M_j, \mathrm{Mexp}_j). \tag{2}$$

Here, $\text{Mexp}_j$ is the measured power distribution at all the output ports when only the input Port $j$ is switched on. It can be accomplished by switching the input channel one by one with the MZIs in the front end of chip. The CF ranges from 0 to 1, where CF=1 or 0 represents that the experimental and theoretical vectors are consistent or irrelevant. Our training target is to make the CF close to 1 as much as possible. The training is accomplished using a modified gradient descent algorithm[41], which is a universal method for training computational neural networks[46-48]. The detailed training algorithm is presented in Supplementary Section 3. In our design, the matrix transformation of $M$ can be implemented by Part (2) or Part (4). We use Part (4) to train the matrix and Part (3) to switch the input channels. Considering that only three input ports ($I_5$, $I_6$, $I_7$) match well to the VGA (where the fiber array has identical spaces), only three-port matrix switches ($I_5$, $I_6$, $I_7$, $O_1$, $O_2$, and $O_3$) are used to carry out the proof-of-concept experiments.

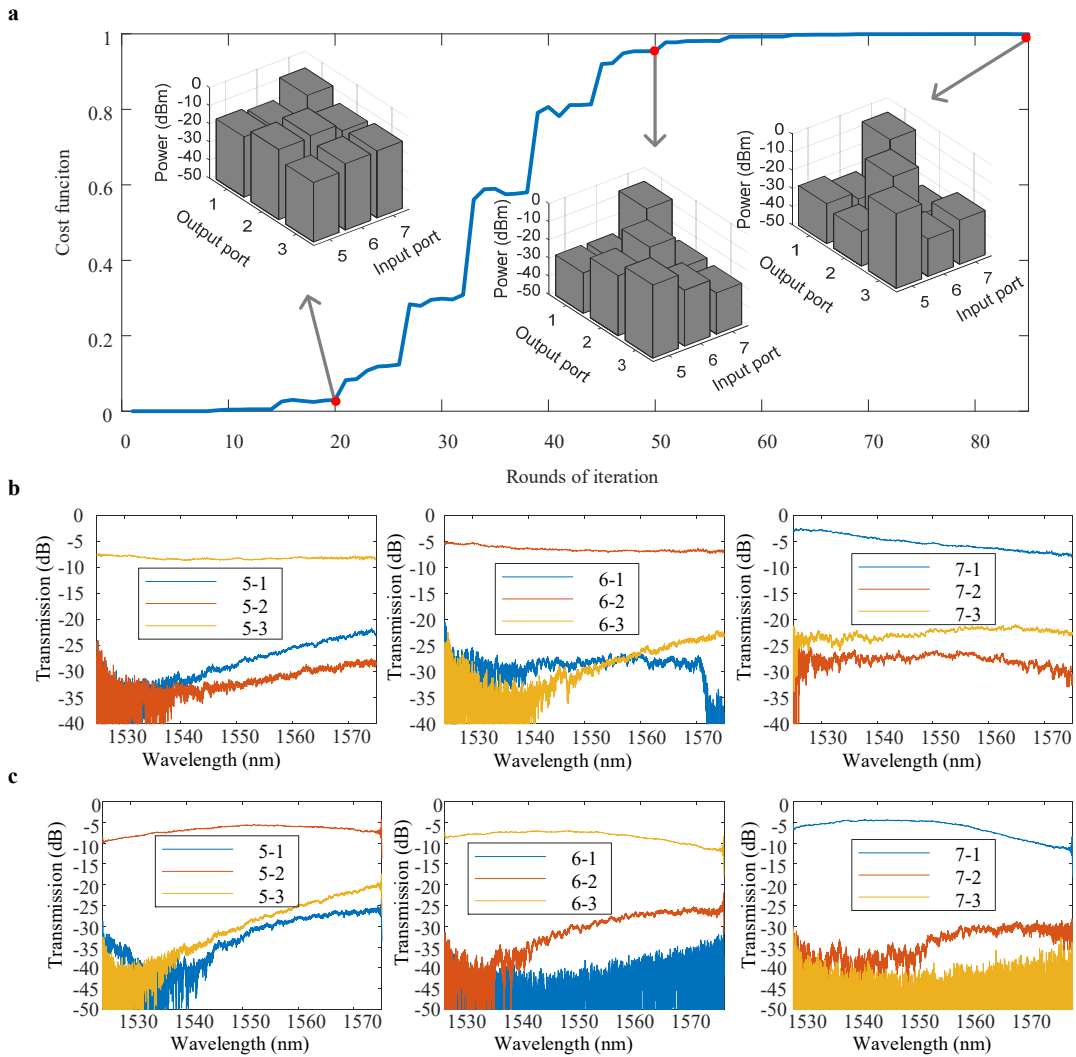

**Fig. 2| Multichannel optical switching.** In the case, only Parts (3) and (4) are used. MZIs in Part (3) are used to switch the input channel, and phase shifters in Part (4) are used to optimize the performance. The input wavelength is set at 1550 nm. **a**, Cost function dependent on the rounds of iteration in training process. The insets figures show the light power distributions when the round of iteration equals 20, 50 and 85 respectively. **b**, Transmission spectra for the signal and noise of the optical links in the shown routing state: $I_5$-$O_3$, $I_6$-$O_2$, $I_7$-$O_1$. **c**, Transmission spectra for the signal and noise of the optical links in the shown routing

state: $I_5$-$O_2$, $I_6$-$O_3$, $I_7$-$O_1$. Also see the supplementary material of Video S1.

To show the reconfigurability and self-learning ability of our chip, we implement multiple routing states for the multichannel optical switching. Figure 2a presents the learning process dependent on rounds of iteration in the routing state: $I_5$-$O_3$, $I_6$-$O_2$, $I_7$-$O_1$ (the iteration evolution is recorded by Video S1). In the beginning, the light power distributions for all output ports are random and messy. And strong crosstalk exists between these channels. The first inset of Fig. 2a shows an example of power distributions when the device has a poor CF less than 0.1, suggesting that the three channels are fully mixed. These channels will be gradually separated with the learning process. The second and third insets in Fig. 2a show the light power distributions when the round of iteration reaches 50 and 85. The crosstalk becomes smaller gradually and the final crosstalk is below -16.8 dB at 1550 nm. Figure 2b shows the measured transmission spectra for different input-output optical routes. The crosstalk in the wavelength range of 1525-1575 nm is below -14 dB. In order to reveal the robustness of our self-learning, we select another routing state of $I_5$-$O_2$, $I_6$-$O_3$, $I_7$-$O_1$ as the training strategy, and the measured transmission spectra are shown in Fig. 2c. The crosstalk values are lower than -23 dB at 1550 nm and lower than -13 dB in the wavelength range of 1525-1575 nm. These results demonstrate the reconfigurability and self-learning ability of the ONN chip for multichannel optical switching. More channels can be configured to switch if the ONN is further expanded.

In fact, the multichannel optical switching can be expanded to other routing networks with multiplexing technology, such as mode-division multiplexing (MDM), space-division multiplexing (SDM), wavelength-division multiplexing (WDM), and polarization-division multiplexing (PDM)[1,5]. For example, the mixed channels distinguished by modes or polarization states can be initially separated and converted to the fundamental modes using the corresponding demultiplexer. Then our routing network is used to reconfigure the routing state using the self-learning method. Finally, a mirrored multiplexer combines these separated channels by converting them to different modes or polarization states. Figure S3 (Supplementary Section 4) depicts an example of mode switch with our multichannel optical switching. Similar set-up is applicable for SDM and PDM channels as well. By combining the demultiplexer and multiplexer, our scheme can be configured for on-chip optical space, mode, wavelength, and polarization switching.

**Optical MIMO descrambler**

The rapidly growing demand for higher transmission capacities promotes the development of MDM and SDM[23,24]. The crosstalk between different channels or modes always exists both in the optical transmission link and in the mode multiplexer/demultiplexer. Traditionally, this crosstalk issue was solved by MIMO algorithms in electronic digital signal processing[49,50], which suffered from heavy computation requirements for high-bandwidth electronic hardware. Alternatively, all-optical MIMO demultiplexing was developed to descramble the modes for all discrete time signals with the inherent speed of light[36,39,51], but these methods were based on manual operation or on the premise of knowing the internal structure of devices. Fortunately, deep learning can be applied to MIMO descrambler as a fully self-learning method. The MIMO descrambler is quite similar to the multichannel optical switching, except that the initial channels are already mixed owing to the crosstalk. In order to recover all channels with low crosstalk, the CF of MIMO descrambler is defined by equation (2), the same with the one of multichannel optical switching. In such a case, the targeted channels are well defined before the crosstalk is introduced. In the experiment, Part (2) is used to emulate the optical transmission link and mode demultiplexer with crosstalk accumulation, thus introducing crosstalk among these channels coming from Part (1). Part (1) is used to switch the input channel. Part (3) and Part (4) are used to eliminate the crosstalk among these channels with the self-learning algorithm.

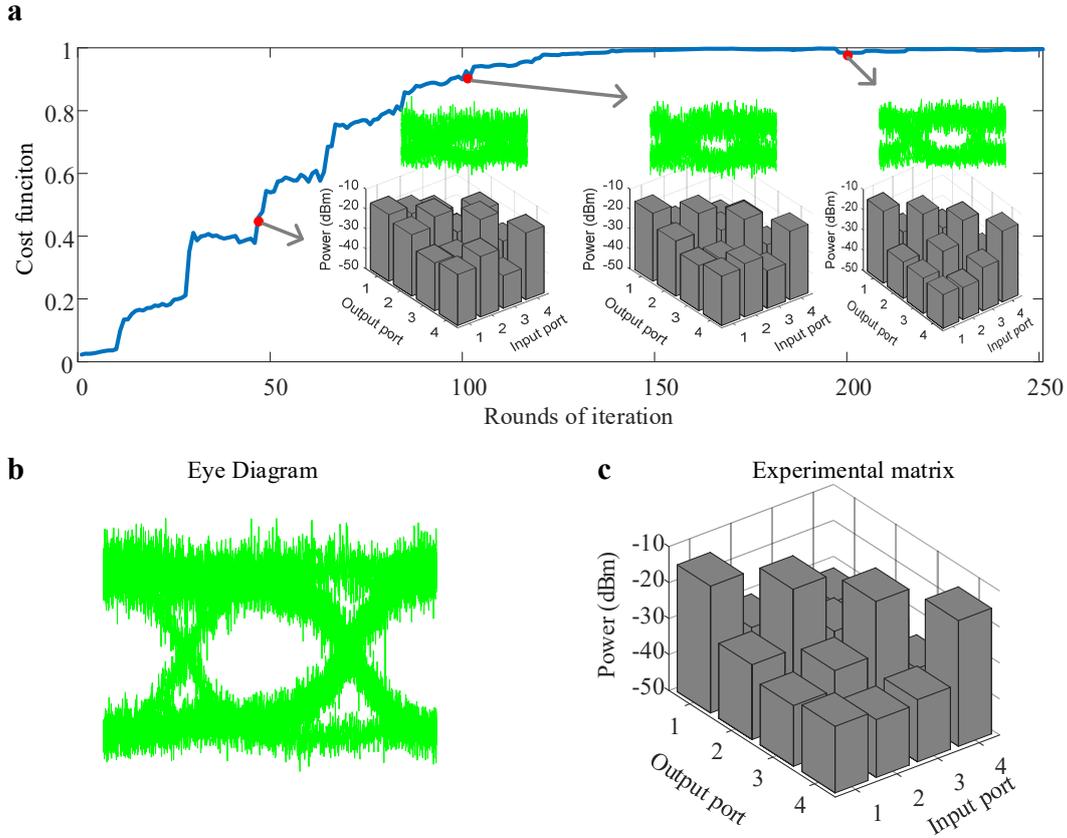

**Fig. 3| Optical MIMO descrambler.** MZIs in Part (1) are used to switch the input channel, and the phase shifters in Part (2) are in random states to add crosstalk among these channels. The phase shifter built in the first MZI of Part (3) is set in the "open" state. All the other phase shifters in Parts (3) and (4) are used to optimize the performance at 1550 nm. **a**, Cost function dependent on the rounds of iteration in training process. The insets figures show the eye diagrams of one monitored channel and light power distributions when the round of iteration equals 50, 100 and 200 respectively. **b**, The final eye diagram of the monitored channel. **c**, The final power distributions in the MIMO structure. Also see the supplementary material of Video S2.

Figure 3a presents the learning process dependent on rounds of iteration in the routing state: $I_1$-$O_1$, $I_2$-$O_2$, $I_3$-$O_3$, $I_4$-$O_4$ (the iteration evolution is recorded by Video S2). Initially, the channel crosstalk is set randomly by applying random voltages on the phase shifters in Part (2). And then the network is trained by tuning the phase shifters in Part (3) and Part (4) to make the CF best. Obviously, the crosstalk is quite strong before the training and the eye starts to open when the round of iteration equals to 50. As the iteration round increases, the CF grows and the crosstalk becomes smaller, which can be seen from the eye diagrams of the monitored channel ($O_1$) and light power distributions when the round of iteration equals to 100 and 200. Finally, the CF almost reaches the best theoretical value of 1. Figures 3b and 3c show the final eye diagram of the monitored channel and light power distribution. The eye is fully open and the crosstalk is lower than -15 dB at 1550 nm. The transmission spectra for all channels are measured and presented in Fig. 4a. The crosstalk is lower than -10 dB in a bandwidth of about 8 nm. Another routing state of $I_1$-$O_3$, $I_2$-$O_2$, $I_3$-$O_1$, $I_4$-$O_4$ is also demonstrated, the measured transmission spectra are shown in Fig. 4b. The crosstalk values are lower than -12 dB at 1550 nm and lower than -10 dB in a bandwidth of about 5 nm. The narrow bandwidth results from compromise on the small crosstalk. Note that in the training process, the input channels are continuously switched and only one channel is switched on in each time. It means that the system cannot propagate light signals for multi-channels in real time during the training process.

In fact, once the MIMO descrambler successfully separates all channels after the training process, the CF can be redefined according to signal quality of receiving channels. For example, we redefine the CF as the opening areas of eye diagrams of receiving channels by

$$\mathrm{CF}=\prod_{j=1}^{n} Sarea_j \tag{3}$$

$Sarea_j$ is the opening area of output eye diagram of Channel *j*. In such a case, the four channels can propagate real-time light signals during the training process. The training strategy is to make the opening areas of eye diagrams in the specified channels maximum. Restrained by the experimental conditions, we train only one channel to demonstrate this modified MIMO descrambler as a proof-of-concept. The evolution of eye diagram is shown in Fig. 5 and recorded by Video S3 respectively. The opening area of eye diagram in the specified channel gradually increases and becomes stable and maximal finally. If multiple channels need to train simultaneously, the CF can be set as the product of opening areas of eye diagrams in targeted channels.

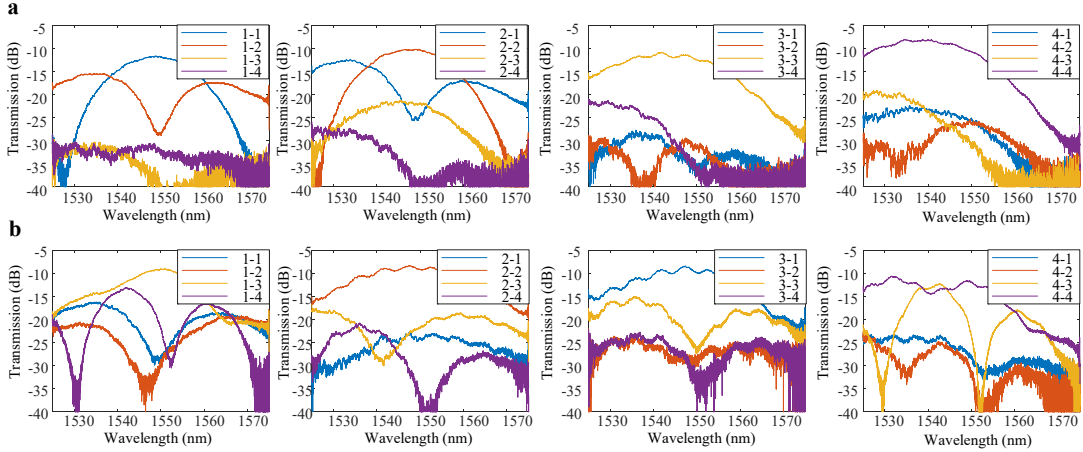

**Fig. 4| Transmission spectra for optical MIMO descrambler. a**, Routing state: $I_1$-$O_1$, $I_2$-$O_2$, $I_3$-$O_3$, $I_4$-$O_4$. **b**, Routing state: $I_1$-$O_3$, $I_2$-$O_2$, $I_3$-$O_1$, $I_4$-$O_4$.

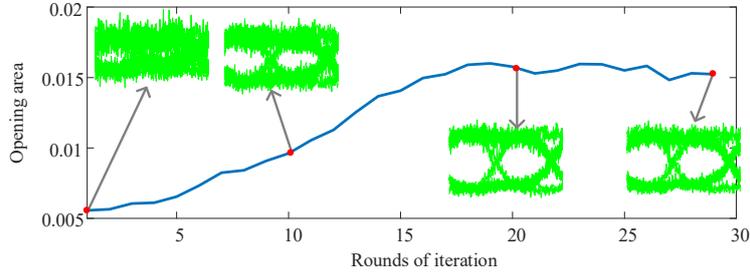

**Fig. 5| Modified MIMO descrambler when all four channels communicate normally.** In the case, all the channels are open and working normally. Alternatively, the signal quality, such as the opening areas of eye diagrams, is used to offer the feedback information to optimize the performance. Also see the supplementary material of Video S3.

**Tunable optical filter**

Optical filter is a basic device in photonic signal processing. The reported tunable optical filters usually offer limited automaticity[12,52]. In our design, the cascaded MZI mesh shows great potential to adjust the transmission spectrum, acting as a smart tunable optical filter. In the experiment, we launch a broadband light source into the chip from input port $I_4$ and an OSA is used to monitor the output spectrum from output port $O_4$. The detailed experimental set-up is presented in Fig. S4 of Supplementary Section 4. The CF is defined as the difference of average powers between the pass band and stop band, given by

$$CF = \overline{P}_{pass} - \overline{P}_{stop} \quad (4)$$

where $\overline{P}_{pass}$ is the average power in the pass band, and $\overline{P}_{stop}$ is the average power in the stop band. All the phase shifters in Parts (2), (3) and (4) are jointly used to optimize the filter. Figure 6a presents the experimental results for tuning the central wavelength from 1537 nm to 1562 nm at a fixed full width at half maximum (FWHM) of 20 nm. Figure 6b presents the measured spectra for bandwidth-tunable filter. The central wavelength is fixed at 1546 nm, and the FWHM is optimized from 18 nm to 32 nm. An example of filter evolution is shown by Video S4. The tunable range is mainly limited by the narrow-band coupling gratings (see Supplementary Fig. S1b).

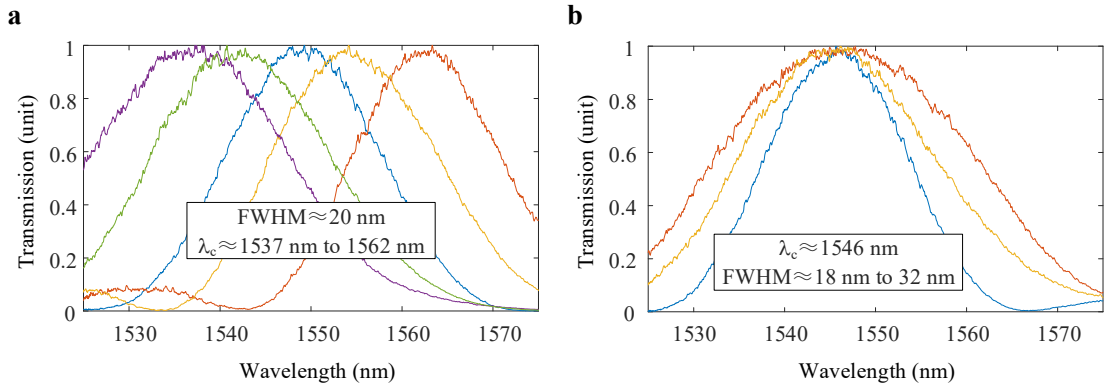

**Fig. 6| Tunable optical filter.** A broadband light source is launched into the chip from $I_4$ and an optical spectrum analyzer is used to monitor the output spectrum from $O_4$, to optimize the filter. All the phase shifters in Parts (2), (3) and (4) are used. **a**, Central wavelength adjustable filter. The full width at half maximum (FWHM) is fixed at 20 nm, and the central wavelength is optimized from 1537 nm to 1562 nm. **b**, Bandwidth adjustable filter. The central wavelength is fixed at 1546 nm, and the FWHM is optimized from 18 nm to 32 nm. Also see the supplementary material of Video S4.

**Discussion and summary**

The proposed photonic signal processor can be reconfigured as multichannel optical switching, optical MIMO descrambler, and tunable optical filter, which are basic building blocks both in the optical communication and optical computing. The proposed photonic signal processor can be used to provide fully self-learning signal processing without knowing the chip inner structure. Multichannel optical switching is one of the essential components in optical communication and interconnect systems. A self-learning optical router can effectively compensate the fabrication errors and avoid the clumsy manual calibration. The proposed two MIMO descramblers can be used in an optical network to perform fast signal processing without electronic digital sampling. They can be combined and applied in MDM-based optical communication. For example, the communication system can be initialized and calibrated using the first MIMO descrambler and then be optimized in real time using the modified MIMO descrambler. The modified MIMO descrambler can work without interrupting the communication, making the communication system has self-healing capabilities in real time. It is significant in optical communication. The tunable optical filter is also demonstrated, but the tunability is limited. It mainly comes from two aspects. On the one hand, the coupling gratings are narrow-band with a 3 dB bandwidth about 40 nm (Supplementary Fig. S1b), thus restraining the tunable range. On the other hand, the MZIs are wavelength-insensitive in the waveband from 1525 nm to 1575 nm, making the tuning sensitivity quite low. A programmable optical filter, such as a shape-tunable filter, can be redesigned by using a more wavelength-sensitive network and a more wide-band coupling method, such as horizontal facet coupling. Our method can learn to configure the targeted functions by itself without knowing the internals of chip, provided the chip has the ability to realize these functions in theory. Thus, the proposed photonic signal processor can provide a potential fully self-learning solution for signal processing in future all-optical networks.

In summary, we have designed, fabricated and demonstrated a reconfigurable and fully self-learning photonic integrated signal processor based on an ONN chip. The signal processor is reconfigured as a multichannel optical switching, two kinds of optical MIMO descramblers, and a tunable optical filter with a tunable central wavelength and a tunable bandwidth. All the functions are programmable by full self-learning starting from blank. The training is accomplished using the numerical gradient descent algorithm modified from deep learning, which is applicable for a "black box" system. More complex and more accurate photonic signal processing functions can be accomplished using a larger network, such as an 8 x 8 ONN. Our demonstration is an important step towards the implementation of a multifunctional optical signal processor with artificial intelligence.

**Methods**

**Device fabrication.** The designed ONN chip is fabricated on the commercial SOI wafer. A passive process is employed to fabricate the structure on silicon-on-insulator (SOI) wafer with 220 nm top silicon and 2 μm thick buried oxide (BOX) substrate. The Si waveguide are 220 nm thick strip waveguide, while the grating coupler is fan-shaped with 70 nm shallow etch. Using 248 nm deep UV lithography, the structure is patterned to photoresist, followed by 70 nm partial grating etch. The remaining part of strip waveguide is then patterned and etched to BOX. Thereafter, Plasma Enhanced Chemical Vapor Deposition (PECVD) is introduced to deposit 1.2 μm pad oxide between Si waveguides and microheaters. Then a layer of 120 nm thick TiN is deposited and etched as the heater layer. Each heater unit is 160 μm long and 1 μm wide. Afterwards, Al is deposited and etched to form the metal wires and pads. The SOI chip is then glued on a ceramic-wafer printed circuit and the pads are wire-bonded to the printed circuit board. The size of chip is about 1.3 mm x 7.5 mm.

**Experimental method.** The ONN chip couples the light with grating couplers connected to VGAs. The pads of chip are wire-bonded to a printed circuit board. The phase shifters are controlled by a VSA and the output power are collected by a PDA, an OSA or an OSC. All the devices are connected and controlled by the same computer. The PDA and VSA are connected to the computer with RS232 serial ports. The OSA and OSC are communicated with Ethernet connections. The computer updates the phase imparted on phase shifters according to the feedback from the monitoring instruments, such as the PDA, OSA and OSC. The targeted functions are configured by full self-learning, without detailed information about the interior structure.

**Acknowledgements**

This work was partially supported by National Natural Science Foundation of China (61622502 and 61805090) and China Postdoctoral Science Foundation (2017M622419).

**Author contributions**

H.L.Z. analyzed the data. H.L.Z., Y.H.Z. and X.W. conceived and designed the experiments. H.L.Z., Y.H.Z., X.W., J.J.D. and X.L.Z. contributed materials and analysis tools. H.L.Z. and Y.H.Z. performed the experiments. H.L.Z., J.J.D. and X.L.Z. wrote the paper.

**Additional information**

Supplementary information is available in the online version of the paper. Correspondence and requests for materials should be addressed to J.J.D.

**Competing financial interests**

The authors declare no competing financial interests.

# Self-learning photonic signal processor with an optical neural network chip

Hailong Zhou[1†], Yuhe Zhao[1†], Xu Wang[1], Dingshan Gao[1], Jianji Dong[1]★ and Xinliang Zhang[1]★

[1]Wuhan National Laboratory for Optoelectronics, School of Optical and Electronic Information, Huazhong University of Science and Technology, Wuhan 430074, China; [†]These authors contributed equally to this work. ★e-mail: jjdong@hust.edu.cn; xlzhang@mail.hust.edu.cn.

## 1. Chip design, fabrication and characterization

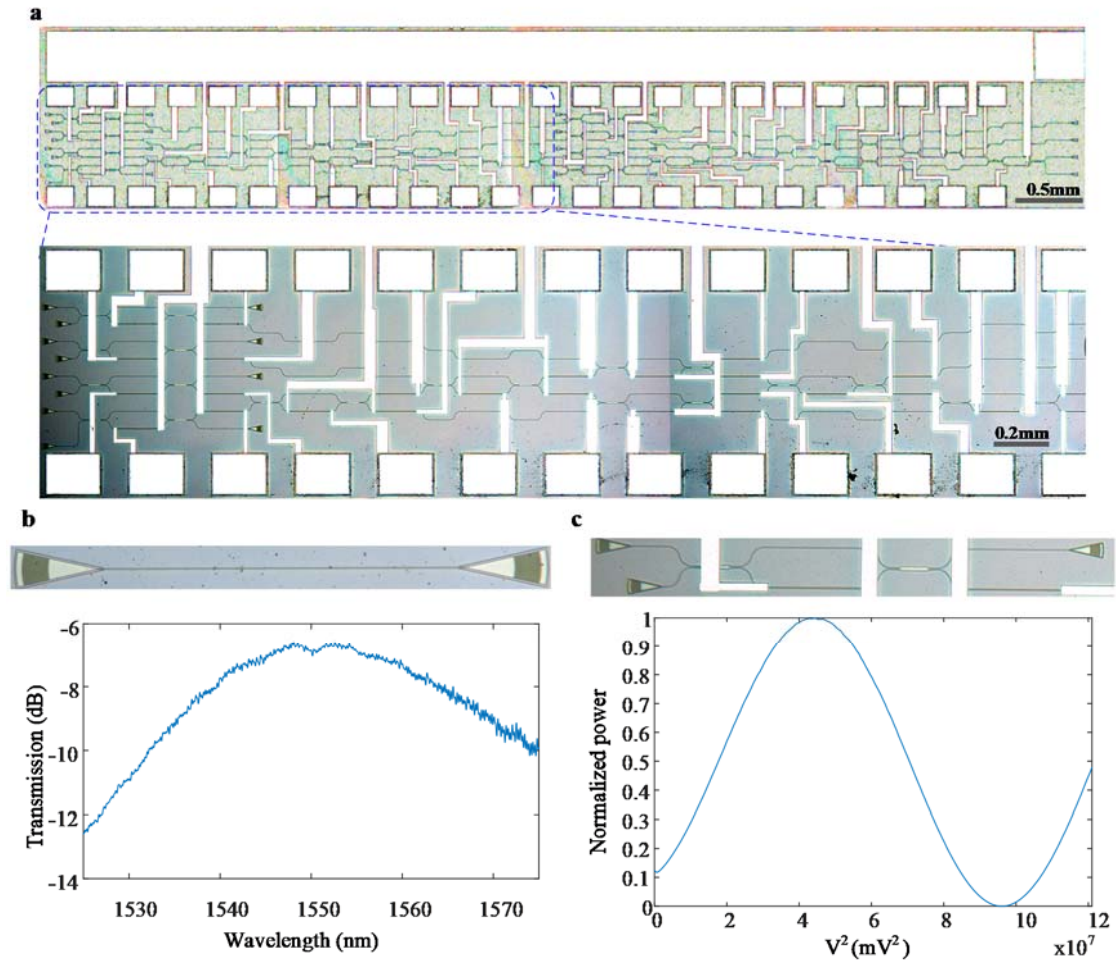

**Fig. S1: Chip fabrication and characterization. a**, The micrographs of fabricated chip. **b**, The transmission spectrum and the used referenced grating structure. **c**, The imparted phase dependent on the applied voltage and the tested MZI structure.

The designed optical neural network (ONN) chip is fabricated on the commercial silicon-on-insulator (SOI) wafer. And the chip size is about 1.3 mm x 7.5 mm. The micrographs of fabricated chip are shown in Fig. S1a. The Mach-Zehnder interferometer (MZI) mesh contains 20 MZIs and 48 phase shifters. Grating coupler arrays are used to couple the light between fibers and the input/output ports. Figure S1b presents the transmission spectrum for the referenced grating structure, the coupling efficiency at 1550 nm is about -6.9 dB and the 3 dB bandwidth is about 40 nm from 1535 nm to 1575nm. Phase shifters introduce a phase change in power dissipated across an electrical resistance to a change in waveguide refractive index by the thermo-optic effect. The imparted phase dependent on the applied voltage ($V$) may be expressed as[1]

$$\theta = \frac{2\pi V^2}{T}. \quad (S1)$$

MZIs with internal phase shifters are used to convert the phase change to the output power. The measured power distribution dependent on the square of applied voltage is presented in Fig. S1c. The measured average period of $T$ is about $1.07 \times 10^8 \, mV^2$. The phase tuning efficiency is measured to be 27 mW per $\pi$ phase shift and the electrical resistance is about 2000 $\Omega$.

## 2. Principle of ONN chip

The nanophotonic processor is programmed by setting the voltages on the internal and external phase shifters of each MZI. The MZI shown in Fig. S2a implements a 2 × 2 unitary transformation on the input state of the form,

$$U_{MZIn} = R(n) = \frac{1}{2} \begin{bmatrix} e^{i\alpha_n}(e^{i\theta_n}-1) & ie^{i\alpha_n}(e^{i\theta_n}+1) \\ ie^{i\beta_n}(e^{i\theta_n}+1) & e^{i\beta_n}(1-e^{i\theta_n}) \end{bmatrix}. \quad (S2)$$

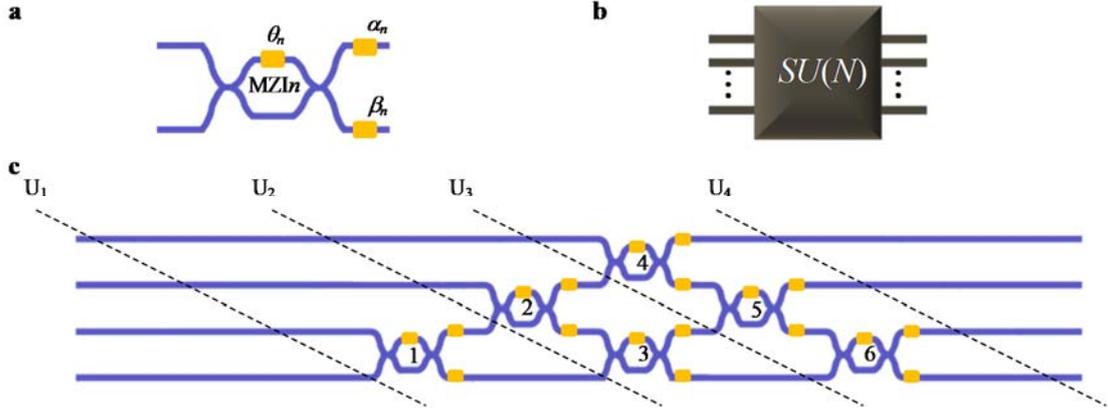

**Fig. S2: Diagram of the nanophotonic processor. a**, Schematic illustration of a single phase shifter in the MZI and the transmission curve for tuning the internal phase shifter. **b**, Sketch map of an *SU(N)* core. **c**, Mesh and MZI Numbering scheme of *SU(4)*

All the phase are normalized to the range of $[0, 2\pi)$. A *SU(N)* core shown in Fig. S2b, namely a unitary matrix with rank *N*, can be decomposed into sets of *SU(2)* rotations implemented by cascaded programmable MZIs[2]. Figure S2c presents an example of *SU(4)* mesh. It contains 6 MZIs and 18 thermo-optic phase shifters. The relationship in different planes can be written by

$$U_2 = R_{1,1}U_1 = \begin{bmatrix} 1 & & \\ & 1 & \\ & & R(1) \end{bmatrix} U_1$$

$$U_3 = R_{2,1}R_{2,2}U_2 = \begin{bmatrix} 1 & & \\ & 1 & \\ & & R(3) \end{bmatrix} \begin{bmatrix} 1 & & \\ & R(2) & \\ & & 1 \end{bmatrix} U_2 \quad (S3)$$

$$U_4 = R_{3,1}R_{3,2}R_{3,3}U_3 = \begin{bmatrix} 1 & & \\ & 1 & \\ & & R(6) \end{bmatrix} \begin{bmatrix} 1 & & \\ & R(5) & \\ & & 1 \end{bmatrix} \begin{bmatrix} R(4) & & \\ & 1 & \\ & & 1 \end{bmatrix} U_3$$

The final $SU(4)$ is given by

$$SU(4) = R_{3,1}R_{3,2}R_{3,3}R_{2,1}R_{2,2}R_{1,1} \quad (S4)$$

The same analysis can be applied on $SU(N)$ core. A unitary matrix with rank $N$ can be generally written as a product of $N(N-1)/2$ rotation matrices

$$SU(N) = R_{N-1,1}R_{N-1,2}\cdots R_{N-1,N-1}\cdots R_{3,1}R_{3,2}R_{3,3}R_{2,1}R_{2,2}R_{1,1}. \quad (S5)$$

To implement any matrix, we first note that a general matrix ($M$) can be decomposed as $M = U\Sigma V$ through singular value decomposition (SVD), where $U$ and $V$ are both unitary matrices, $\Sigma$ is a diagonal matrix with its eigenvalues on the diagonal. In our design, we use the structure in Fig. 1a to implement any 4 x 4 matrix.

## 3. Training of multiple functions with ONN chip

### 3.1 Defining CF of targeted functions

For certain targeted function, the cost function (CF) is vital for a successful training. All the training will proceed to make the CF best. In our experiment, four sub-experiments are carried out with different CF definition. The multichannel optical switching and optical multiple-input-multiple-Output (MIMO) descrambler share the same CF. For these two functions, the targeted transmission matrix is defined by

$$M = \begin{bmatrix} M_1 & M_2 & \cdots & M_n \end{bmatrix} = \begin{bmatrix} m_{11} & m_{12} & \cdots & m_{1n} \\ m_{21} & m_{22} & \cdots & m_{2n} \\ \vdots & \vdots & \vdots & \vdots \\ m_{n1} & m_{n2} & \cdots & m_{nn} \end{bmatrix} \quad (S6)$$

where $|m_{ij}|=0,1$ indicates the optical routing state. $|m_{ij}|=1$ represents the light launched into input Port $j$ will emerge from output Port $i$. $n$ is the number of channels. The matrix can be measured experimentally by switching the input channel one by one, for example, the output power distribution collected by the photodetector array (PDA) will equal to $M_j$ when a unit light is launched into input Port $j$. Generally, a correlation function can be used to characterize the level of similarity between two vectors, given by

$$\text{Corr}(a,b) = \frac{|a \cdot b|}{|a||b|}, \quad (S7)$$

The operation '•' means the scalar product of two vectors. Considering the unbalanced input power in our design, our CF is defined channel-independently by

$$\text{CF} = \prod_{j=1}^{n} \text{Corr}(M_j, \text{Mexp}_j), \quad (S8)$$

Here $\text{Mexp}_j$ is the measured output power distribution when only the input Port $j$ is open. In the training process, the input channels are continuously switched and only one channel is open in each moment. It means that the system cannot work until the training is complete. In fact, this issue can be solved by redesigning the CF. In the case, all the channels are switched on, and the CF is designed only according to signal quality of receiving channels, therefore the system can work normally in the training process. Here, we define the CF as the opening areas of eye diagrams of receiving channels during the MIMO training, which is defined by

$$\text{CF}=\prod_{j=1}^{n} Sarea_j, \quad (S9)$$

$Sarea_j$ is the opening area of output eye diagram of Channel $j$. In the experiment, the CF can be set as the product of opening areas of eye diagrams of one or more targeted channels.

For the tunable optical filter, only the ports $I_4$ and $O_4$ are used, the CF is defined as the difference of average powers between the pass band and stop band, given by

$$\text{CF}=\overline{P}_{pass}-\overline{P}_{stop}, \quad (S10)$$

where $\overline{P}_{pass}$ is the average power in the pass band, and $\overline{P}_{stop}$ is the average power in the stop band.

**3.2 Numerical Implementation**

Theoretically, the training needs to combine forward propagation and backward propagation methods similar to the deep learning[1]. The forward propagation is used to calculate the output as the data for the next iteration and then the backward propagation aims to estimate the errors and find the gradient descent. This training algorithm is also called by gradient descent algorithm, which is a common method for training artificial neural networks (ANNs). In our design, the ONN chip can output automatically and timely, provided the input is set. And the gradient descent can be alternatively measured by fine tuning each parameter. So in our design, no backward propagation is needed and the forward propagation can be finished by the ONN itself at the speed of light. The main cost of time is the update time of electronic control signal provided by external driver and the response time of the phase shifters in the ONN chip. Furthermore, the ONN chip can be regarded as a "black box", the internal structure of the chip is sightless for training. The full training process is listed below:

1. Initialization: all the adjustable parameters of $\theta_i (i=1,2,...)$ are set randomly.

2. Tuning each parameter: set $\theta_1$ to $\theta_1+\Delta\theta$ temporarily.

   If $\text{CF}(\theta_1+\Delta\theta) \geq \text{CF}(\theta_1)$, replace $\theta_1$ with $\theta_1+\Delta\theta$;

   Else, replace $\theta_1$ with $\theta_1-\Delta\theta$.

3. Repeat Step 2 for all adjustable parameters one by one.
4. Repeat Steps 2 and 3 until the CF is converged or reach target value.

The step size of $d\theta$ can be changed in the training process, for example, we use the step size of $d\theta=2\pi/20$ to find a rough solution firstly and then use step size of $d\theta=2\pi/100$ to further optimize the solution. And in order guarantee the power efficiency, the first eigenvalue is fixed and equals 1, namely, the first MZI in diagonal matrix of Part (3) is always open in the training process.

**4 Supplementary figures**

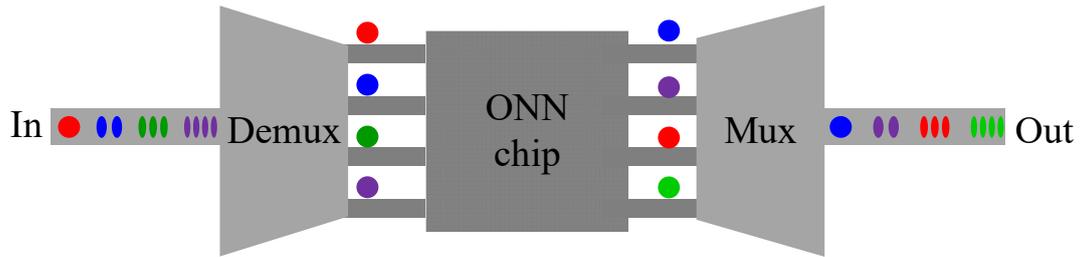

**Fig. S3: Schematic of the set-up for multichannel optical switching based on modes.** The mixed channels distinguished by modes are initially separated and converted to the fundamental modes using the corresponding demultiplexer. Then the ONN chip is used to reconfigure the routing state using the self-learning method. Finally, a mirrored multiplexer combines these separated channels and reconverts them to different modes. Demux, demultiplexer; Mux, multiplexer. Similar set-up is applicative for SDM and PM.

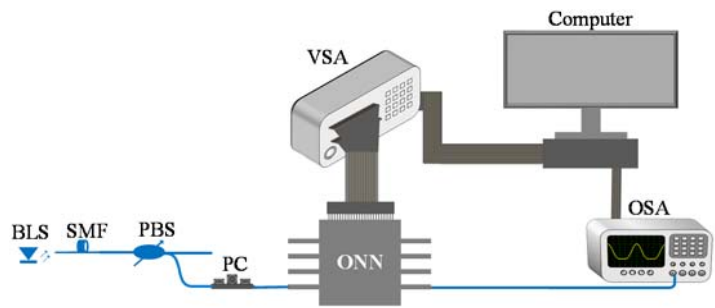

**Fig. S4: Schematic of the experimental set-up for tunable optical filter.** A broadband light source (BLS) is launched into the chip from $I_4$. The polarization beam splitter (PBS) and polarization controller (PC), connected by single mode fibers (SMFs), are used to tune polarization states to ensure coupling efficiency. An optical spectrum analyzer (OSA) is used to monitor the output spectrum from $O_4$, to optimize the filter. The chip is controlled by a voltage source array (VSA). The VSA and OSA are connected and controlled by the same computer.